\documentclass[%
reprint, 
superscriptaddress, 
 amsmath,amssymb,
 aps,
]{revtex4-2}

\usepackage[colorlinks=true, pdfstartview=FitV, linkcolor=black, citecolor=black, urlcolor=black]{hyperref}
\usepackage{graphicx,color}
\usepackage{dcolumn}
\usepackage{bm}
\usepackage{float,hyperref}
\usepackage{orcidlink,lineno}

\begin{document}

\preprint{APS/123-QED}

\title{Replication Time-Lag–Induced Lethality in Quasispecies}
\title{Quasispecies Lethality Driven by Replicative Time Lags}
\title{Replication Time Lags as a Novel Route to Viral Quasispecies Extinction}

\title{Lag-driven Dynamical Transitions as a Novel Route to Virus  Quasispecies Extinction}

\title{Lag-Induced Dynamical Transitions as a Route to Quasispecies Extinction}

\title{Lag-Induced Critical Transitions to Extinction in Replicating Systems}

\author{Edward~A.~Turner\orcidlink{0000-0002-2959-9227}}
\email{Corresponding author: edward.turner@uvm.cl}
\affiliation{Facultad de Ciencias Jurídicas, Sociales y de la Educación, Universidad Viña del Mar. Diego Portales 90, Viña del Mar, Chile.}
\author{Francisco~Crespo\orcidlink{0000-0002-5930-8523}}
\affiliation{Aerospace Engineering Department, Embry-Riddle Aeronautical University. Daytona Beach, FL 32114, USA.}
\author{Joan Gimeno\orcidlink{0000-0002-8707-6379}}
\affiliation{Departament de Matemàtiques i Informàtica, Universitat de Barcelona (UB). Gran Via de les Corts Catalanes 585, 08007 Barcelona, Catalonia, Spain}
\author{Ernest~Fontich\orcidlink{0000-0002-2415-9310}}
\affiliation{Departament de Matemàtiques i Informàtica, Universitat de Barcelona (UB). Gran Via de les Corts Catalanes 585, 08007 Barcelona, Catalonia, Spain}
\affiliation{Centre de Recerca Matem\`atica (CRM). Edifici C, Campus de Bellaterra 08193 Cerdanyola del Vallès, Barcelona, Spain}
\author{Santiago~F.~Elena\orcidlink{000-0001-8249-5593}}
\affiliation{Institute for Integrative Systems Biology (I$^2\!$SysBio), 
CSIC-Universitat de València, Av. Catedràtic Agustín Escardino Benlloch 9, Paterna, 46980 València, Spain}
\affiliation{The Santa Fe Institute, 1399 Hyde Park Road, Santa Fe, NM87501, USA.}
\author{Josep~Sardanyés\orcidlink{0000-0001-7225-5158}}
\email{Corresponding author: jsardanyes@crm.cat}
\affiliation{Centre de Recerca Matem\`atica (CRM). Edifici C, Campus de Bellaterra 08193 Cerdanyola del Vallès, Barcelona, Spain}
\affiliation{Dynamical Systems and Computational
Virology, CSIC Associated Unit I$^2\!$SysBio-CRM, Spain.}



\begin{abstract}
Replicating systems sustained by error-prone enzymatic amplification can undergo critical transitions between persistence and extinction. In RNA viruses, such transitions are classically governed by mutation rates and fitness landscapes, giving rise to error thresholds and lethal mutagenesis. Motivated by experimental evidence that polymerase-targeting antivirals constrain replication, we analyze replicating systems with explicit delays in replication-enzyme availability. We identify a lag-induced (dynamical) critical transition driven by the loss of temporal coordination between genome translation and replication. At a fixed mutation rate and replicative fitness landscape, populations cross an extinction threshold solely due to time delays. Within the quasispecies framework, replication–translation timing emerges as an independent control parameter, defining a distinct dynamical route to extinction and suggesting new antiviral strategies based on modulating replicase availability. More generally, we propose that the pathway to collapse described in this article can be understood as lag-time-induced tipping ($\tau$-tipping).

\end{abstract}

\maketitle

\begin{figure}
\centering
\includegraphics[width=\columnwidth]{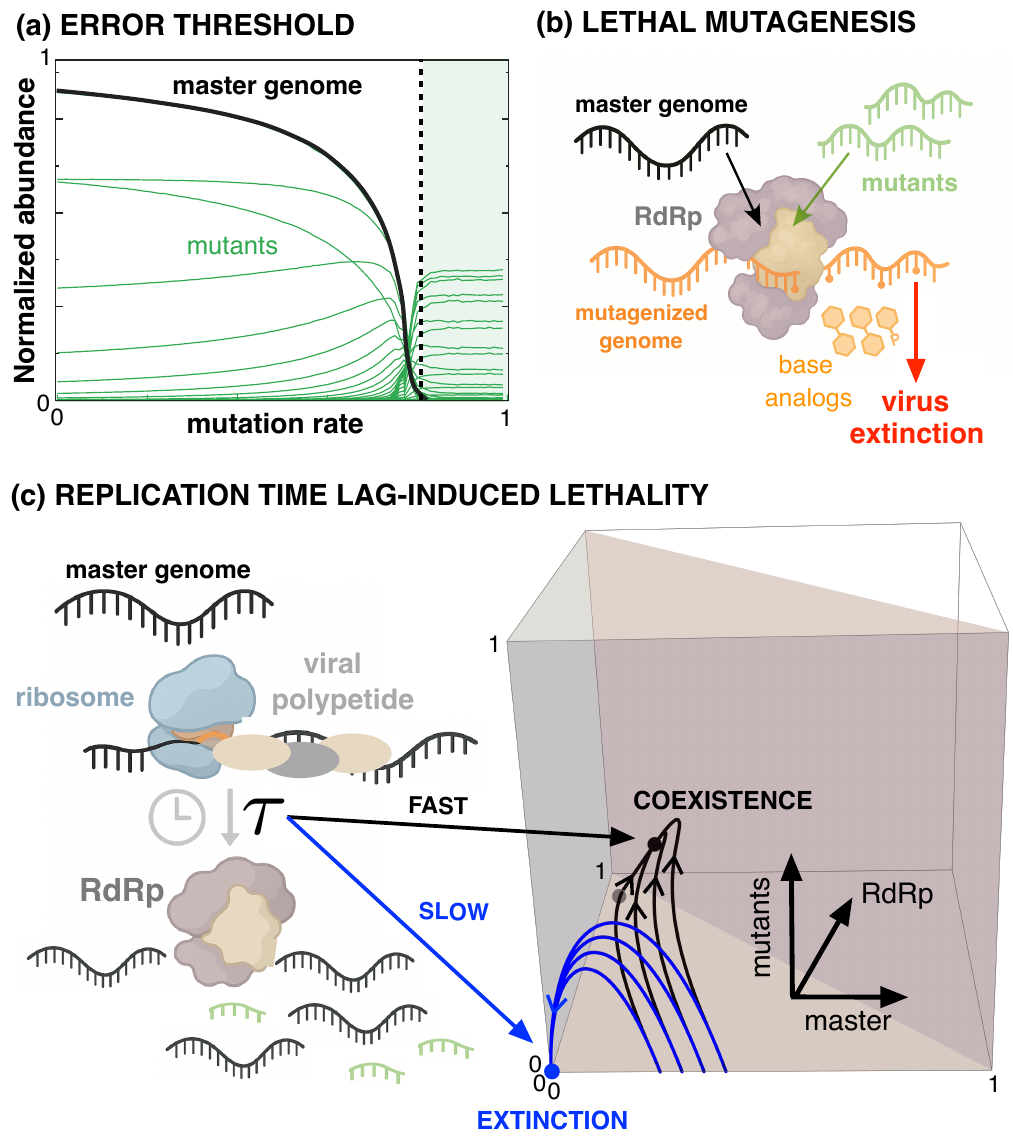}
\caption{Critical transitions in error-prone self-replicating population, such as RNA viruses. (a) The error threshold marks mutant dominance over the master genome~\cite{Eigen2002,Sole2006}. (b) Lethal mutagenesis causes extinction via mutation accumulation~\cite{Loeb1999,Bull2007}. (c) Replication time--lag-induced lethality identified in this study. Our model tracks master and mutant genomes replicated by the RNA-dependent RNA polymerase (RdRp), with a time lag $\tau$ in its functional availability. Phase-space trajectories are computed from Eqs.~\eqref{eq:DIviruses} for different situations: early RdRp production ($\tau=0.65$, black) supports persistence, whereas long lags ($\tau=50$, blue) cause collapse. Panels (a,b) show mutation- and fitness-driven transitions, while (c) shows extinction driven solely by replication timing at fixed mutation rate and replication fitness landscape.}
\label{fig1}
\end{figure}
{\it Introduction}\textemdash 
Replicating populations are nonequilibrium systems that can undergo sharp transitions between persistence and extinction. Classical theoretical approaches have emphasized mutation, selection, and replication fitness landscapes as the primary determinants of these transitions. Within this setting, quasispecies theory provides a canonical framework for describing the collective dynamics of replicating genomes under high mutation rates \cite{Eigen1971,Eigen1989,Domingo2000}, with RNA viruses as a prototypical realization. A quasispecies is a genetically heterogeneous population maintained around a master sequence, whose dynamics can exhibit critical transitions in information fidelity and population survival. Several extinction thresholds have been identified, including the error threshold \cite{Eigen1971,Eigen1989}, lethal mutagenesis \cite{Loeb1999,Bull2007}, and lethal defection \cite{Domingo2018}. Beyond their theoretical significance, these transitions underpin antiviral strategies aimed at driving viral populations to extinction \cite{Eigen2002,Domingo2019,Peck2018}.

Experimental and computational studies have provided evidence for distinct mutational transitions across RNA viruses. Error catastrophe, associated with the loss of genetic information beyond a mutation threshold, was first demonstrated in poliovirus treated with ribavirin~\cite{Crotty2001} and later inferred in hepatitis~C virus (HCV) data using quasispecies models~\cite{Sole2006}, with computational studies suggesting that human immunodeficiency virus type-1 (HIV-1) operates near this threshold~\cite{Tripathi2012}.  Lethal defection, driven by the accumulation of defective interfering genomes, has been observed in lymphocytic choriomeningitis virus and tobacco mosaic virus under fluorouracil treatment~\cite{Grande2005,Diaz2018}. Lethal mutagenesis, characterized by extinction through mutation accumulation under mutagenic pressure, has been reported in foot-and-mouth disease virus, HCV, HIV-1, influenza~A virus, and, more recently, SARS-CoV-2~\cite{Loeb1999,Sierra2007,Ortega2013,Baranovich2013,Kabinger2021}. Together, these results demonstrate the broad relevance of informational and population-level thresholds in viral evolution and pathogenesis.


All previously characterized routes to viral quasispecies extinction rely on changes in parameters such as mutation rates, replication efficiencies, or competitive asymmetries between genomes. Here, we show that extinction can instead arise solely from temporal constraints in replication, even when mutation rates and the replication fitness landscape remain unchanged. This identifies replication timing, relative to protein translation, as an independent dimension of control in quasispecies dynamics, defining a class of critical transitions toward extinction that cannot be mapped onto existing mechanisms. In particular, the virus-encoded RNA-dependent RNA polymerase (RdRp), responsible for viral genome amplification, must first be translated from viral messenger RNAs by host ribosomes and acquire its proper folding and localization before becoming functional. Together with these steps, additional processes can introduce further time lags in replication dynamics (see Appendix 1). 

In this Letter, we describe an alternative route to extinction arising from replicative time lags ($\tau$) between genome replication and protein translation, with particular emphasis on the availability of the viral RdRp. We analyze a minimal dynamical model describing a replicating population on a single-peak fitness landscape, consisting of a master sequence and an average mutant class \cite{Bull2005,Sole2006}. Enzymatic replication is explicitly incorporated by treating RdRp as a dynamical variable, whose production and functional activation occur with finite time delays. 
These replicative lags, able to induce quasispecies extinction, may be considered as a novel control parameter in quasispecies dynamics, extending the theoretical framework of error-driven transitions and suggesting new paths in antiviral intervention. 

{\it Replication lag model}\textemdash We consider a dynamical system derived from the Eigen--Schuster quasispecies model~\cite{Eigen1971}, in which the population fractions $x_i$ of $n$ replicating genomes evolve according to the differential equation $\dot{x}_i=\sum_{j=1}^n x_j f_j Q_{ij}-\phi(\mathbf{x})x_i,$ where $f_j$ denotes the replication rate of genome $j$, $Q_{ij}$ the mutation probabilities, and $\phi(\mathbf{x})$ the mean population fitness. In a minimal reduction~\cite{Swetina1982}, we retain two effective classes: the master sequence $x_0$ and an aggregated mutant class $x_1$, assuming mutations occur from the master to the mutant compartment but not \emph{vice versa}, a standard approximation justified by the large sequence space~\cite{Sole2003}. Unlike the original Eigen--Schuster formulation, we allow the total population to vary in time, reflecting replication from a small initial pool of genomes. Crucially, we explicitly include the viral RdRp, denoted by $p$, as a dynamical variable. This leads to the following set of delay differential equations (DDEs), which incorporate finite replication lags:
{\small
\begin{equation}
\begin{split}\label{eq:DIviruses}
 \frac{dx_0(t)}{dt} &= (1 - \mu) \gamma x_0(t)\ r_0  p(t-\tau) \Phi(x) - \varepsilon_0\,  x_0(t),\\
 \frac{dx_1(t)}{dt}  &= p(t-\tau) \left[\mu \gamma\, r_0 x_0(t) + (1-\gamma) \,r_1 \,x_1(t) \right] \Phi(x) - \varepsilon_1 x_1(t),\\
 \frac{dp(t)}{dt}  &= k \, x_0(t) \left[1 - p(t)\right] - \varepsilon_p \, p(t),
\end{split}
\end{equation}
}
\begin{figure}[b]
\centering
\includegraphics[width=\columnwidth]{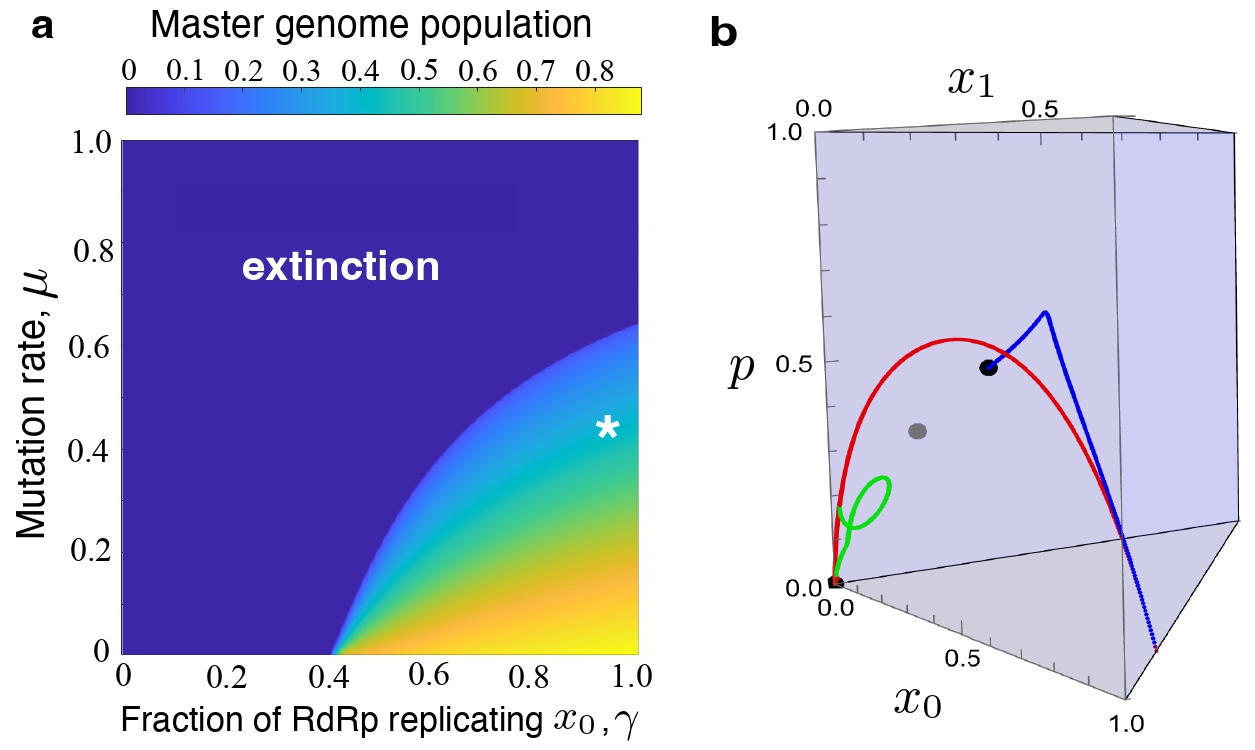}
\caption{(a) Phase diagram of the master genome equilibrium of eqs.~\eqref{eq:DIviruses} for $\tau=0$, showing persistence at low $\mu$ and large $\gamma$. The blue area corresponds to the full collapse of the system. (b) Phase portrait at $(\gamma,\mu)=(0.9,0.4)$ [asterisk in (a)] showing three trajectories with same initial condition for $\tau=0$ (blue), $\tau=25$ (red trajectory overlapped to the green one), and $\tau = 50$ (green). Increasing $\tau$ drives extinction. Black and gray dots denote stable equilibria and saddles.}
\label{fig2:diagram}
\end{figure}
where $\Phi(x)=1-x_0(t)-x_1(t)$ accounts for competition between genomes under a normalized carrying capacity. Mutants are modeled as defective in a functional RdRp  and replicating by complementation with the master's RdRp. Following~\cite{Sardanyes2010,Llopis2025}, RdRp production is proportional to the abundance of master genomes and bounded by logistic growth. Master and mutant genomes replicate at rates $r_0$ and $r_1$, and decay at rates $\varepsilon_0$ and $\varepsilon_1$, respectively. Viral replication is proportional to RdRp abundance, which becomes functional after a time lag $\tau$, representing a causal constraint associated with RdRp synthesis and maturation that cannot be captured by instantaneous models. This delay introduces a de-synchronization between the translation of the viral genomes that first infect the cell into RdRp and their subsequent maturation and the moment in which these recently synthesized RdRp start producing the progeny of viral genomes. A fraction $\gamma$ of RdRp replicates master genomes, with mutation occurring at rate $\mu$. For simplicity, we set $r_0=1$, $r_1<r_0$, and $\varepsilon_0=\varepsilon_1=\varepsilon_p=\varepsilon$, a choice that preserves the qualitative dynamics while enabling analytical tractability.

{\it Lag-induced dynamical transition}\textemdash To assess the impact of time lags in RdRp synthesis and maturation, we first consider the instantaneous case ($\tau=0$). Equations~\eqref{eq:DIviruses} admit three equilibria (see Appendix~2), leading to bistability: the locally stable origin ($P_0$); a stable coexistence equilibrium ($P_+$) involving all three variables; and an interior saddle point ($P_-$). Figure~\ref{fig2:diagram}(a) shows the phase diagram of the equilibrium master-genome populations derived from $x_{0+}$ as functions of $\gamma$ and $\mu$ at $\tau=0$. The boundary between persistence and extinction of the master genome corresponds to a saddle--node bifurcation; extinction of both the master genome and mutants occurs for increasing mutation rates or decreasing fractions of RdRp replicating the master genome.

\begin{figure}
\centering
\includegraphics[width=\columnwidth]{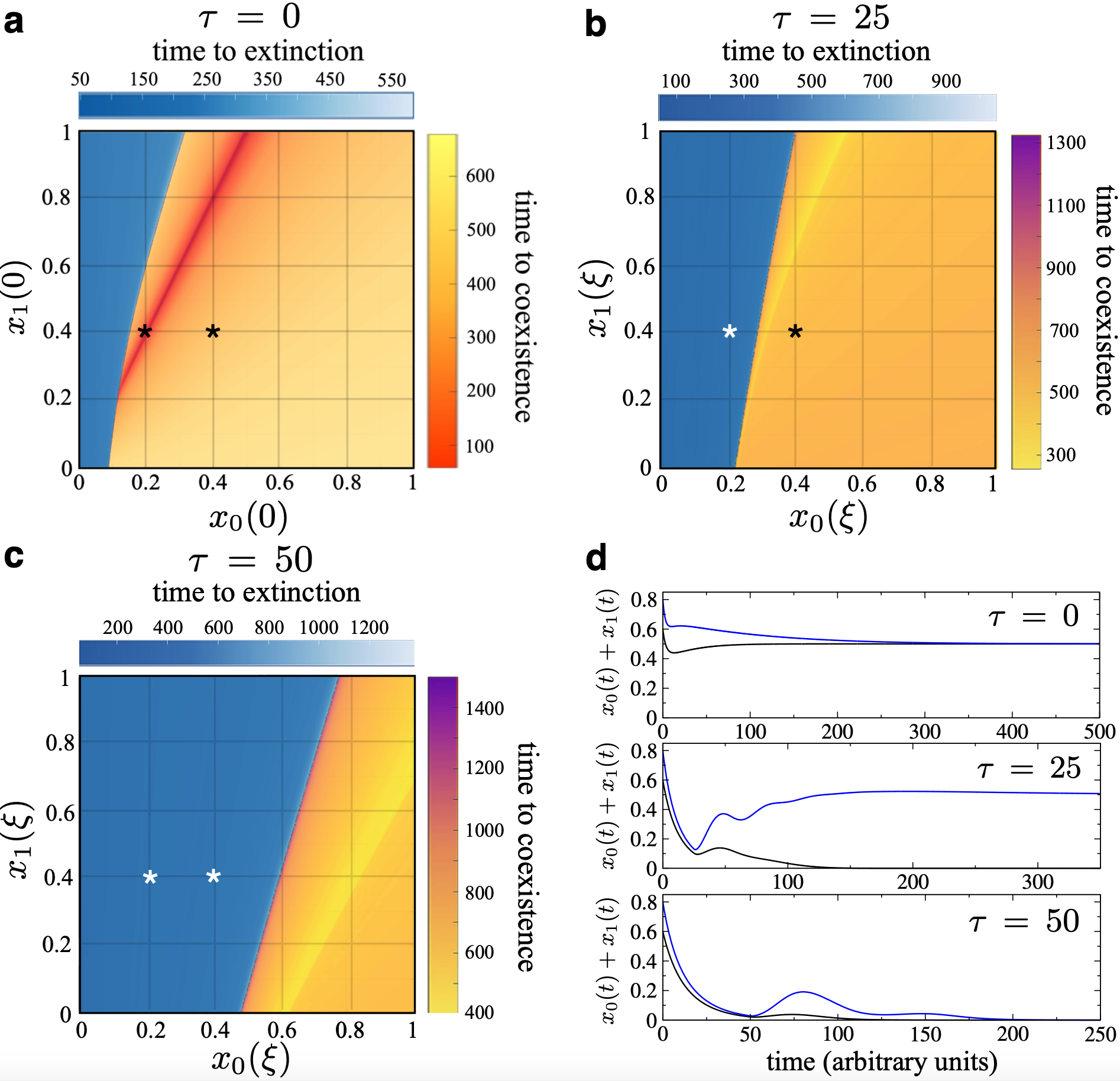}
\caption{(a–c) Basins of attraction of eqs.~\eqref{eq:DIviruses} for $p(0)=0.1$ with $\gamma = 0.5$ and $\mu=0.2$, increasing $\tau=0$, $25$, and $50$, showing convergence times to extinction (blue) or to persistence (orange) for $\xi\in[-\tau,0]$. Increasing $\tau$ enlarges the extinction basin without changing mutation or fitness traits. (d) Time series from the asterisked initial conditions.}
\label{fig3:basins}
\end{figure}

Numerical integration of eqs.~(2)~\footnote{Numerical integration of the DDEs has been performed with the Dormand–Prince Runge–Kutta scheme (order 4/5) using the method of steps.} confirms the analytical predictions. For $\gamma=0.9$ and $\mu=0.4$ [Fig.~\ref{fig2:diagram}(a)], trajectories converge to the persistence equilibrium. Increasing the replication time lag $\tau$ redirects the same initial conditions toward extinction, without changing mutation rates or replication fitness. This transition cannot be reproduced by any effective parameter renormalization and is therefore intrinsically time-driven. Figure~\ref{fig2:diagram}(b) illustrates persistence for $\tau=0$ (blue trajectory) and extinction for larger delays (red and green orbits). As a result, increasing $\tau$ reshapes the basins of attraction, markedly expanding the basin of the extinction state, as quantified in Fig.~\ref{fig3:basins}.

To clarify how changes in $\tau$ alter the basins of attraction, we next consider a simplified model that shows the same phenomenon.

{\it Minimal lag model}\textemdash To gain insight into the basin restructuring induced by large $\tau$, we introduce a minimal two-variable model that reproduces the qualitative behavior of eqs.~\eqref{eq:DIviruses}, consisting of a replicator $x$ that encodes its polymerase $y$ with logistic growth and linear degradation. The model reads
\begin{equation}
\begin{split}\label{eq:simpler}
 \frac{dx(t)}{dt} &= x(t) \,y(t-\tau)\, [1-x(t)] - \varepsilon\,  x(t),\\
 \frac{dy(t)}{dt}  &= x(t)\, [1 - y(t)] - \varepsilon \, y(t).
\end{split}
\end{equation}
The system has three equilibrium points given by the origin $P_0 = (0,0)$, and the pair $P_{\pm} = \left( x_{\pm},\, x_{\pm}/(x_{\pm} + \varepsilon) \right)$, with $x_{\pm} = [(1 - \varepsilon) \pm \sqrt{1 - 2\varepsilon - 3\varepsilon^2}]/2$. The points $P_{\pm}$  exist for $0 \le \varepsilon \le 1/3$. Linear stability analysis shows that the origin has eigenvalues $\lambda_{1,2} = -\varepsilon$ and is therefore locally stable. The evaluation of the Jacobian shows that $\det J(P_-) < 0$, thus indicating that $P_-$ is a saddle point, while $\det J(P_+) > 0$ and $\operatorname{tr}J(P+) < 0$, showing that $P_+$ is a stable node. These two latter equilibria merge at $\mathbf{x}_c = (1/3,\, 1/2)$ in a saddle-node bifurcation when $\varepsilon = \varepsilon_c = 1/3$, $\varepsilon_c$ being the bifurcation value. 

\begin{figure*}
\centering
\includegraphics[width=\textwidth]{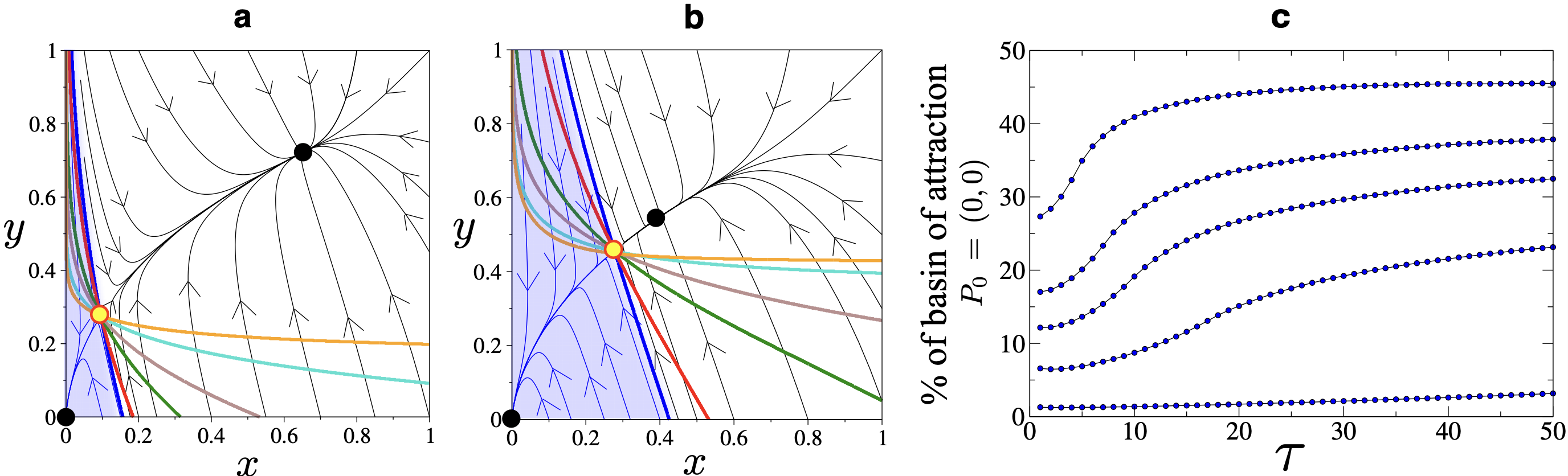}
\caption{Phase portraits of Eqs.~\eqref{eq:simpler} for $\tau=0$ with $\varepsilon=0.25$ (a) and $\varepsilon=0.33$ (b). Trajectories converge to $P_0$ or $P_+$; the basin of $P_0$ is shown in transparent blue (black circles denote attractors; yellow circle denote the saddle point $P_-$). Overlapped, we display the separatrix of the two attractors for different time lags, with: $\tau = 0$ (blue curve); $\tau = 1$ (red curve); $\tau = 5$ (green curve); $\tau = 10$ (brown curve); $\tau = 25$ (turquoise curve); and $\tau = 50$ (orange curve). (c) Basin size of $P_0$ versus $\tau$ for $\varepsilon=0.15$, $0.25$, $0.29$, $0.31,$ and $0.33$ (bottom to top).}
\label{fig4:simple_model}
\end{figure*}

This simplified model allows an analytical characterization of the lag-induced basin reorganization through the $\tau$-dependent spectrum of nontrivial equilibria (Appendix~3). We next investigate the resulting dynamics numerically.  Fig.~\ref{fig4:simple_model} illustrates the dynamics of the simplified system~\eqref{eq:simpler} through phase portraits for two values of the degradation rate $\varepsilon$ at $\tau=0$. For low $\varepsilon$ [panel (a)], the stable coexistence equilibrium $P_+$ lies far from the saddle $P_-$, whose stable manifold defines the separatrix between extinction and persistence. The basin of attraction of the extinction state $P_0$ is shown in blue. Superimposed are numerically computed separatrixes for increasing replication delays ($\tau=1,5,10,25,50$), which progressively shift toward larger $x$, indicating a marked expansion of the extinction basin along the genome-density direction. A similar delay-induced reorganization is observed for larger $\varepsilon$ [panel (b)], where $P_+$ approaches the saddle and the extinction basin again expands with $\tau$. Under biologically relevant initial histories, $x(\xi)\ll1$ and $y(\xi)=0$ for $\xi\in[-\tau,0]$, increasing replication time lags, therefore, bias trajectories toward extinction, consistent with the full model.

Finally, we quantify the expansion of the extinction basin induced by the replication lag $\tau$. For different $\varepsilon$, the lagged system was integrated from a $150\times150$ uniform grid in $(x, y) \in[0,1]\times[0,1]$, and trajectories were classified by convergence to extinction or persistence. The resulting basin fractions are shown in Fig.~\ref{fig4:simple_model} for $\varepsilon=0.15, 0.25,0.29, 0.31, 0.33$. In all cases, increasing $\tau$ rapidly enlarges the extinction basin, with a greater widening effect as $\varepsilon\to\varepsilon_c$ (Appendix~4).

Several experimental studies suggest that viral extinction can be strongly influenced by externally modulating the availability of the replication machinery. In particular, antiviral inhibitors targeting viral polymerases reduce effective replilcation and viral load, and can synergize with mutagenic treatments to drive RNA virus populations to extinction \cite{Sierra2000,Pariente2001}. While these works were interpreted within mutation- and fitness-based frameworks, they are also consistent with a dynamical scenario in which limiting RdRp availability introduces effective temporal constraints on replication, providing empirical motivation for lag-driven extinction mechanisms.

Although motivated by RNA virus replication, the mechanism described here is generic to autocatalytic or crosscatalytic systems, \emph{i.e.}, hypercycles, with lagged catalytic activity. From a dynamical-systems perspective, time lags are known to reshape basins of attraction and alter attractor selection without changing instantaneous control parameters~\cite{YanWiercigroch2017,Leng2016,PisarchikFeudel2014}. In this framework, the extinction reported here corresponds to a time-lag–driven reorganization of basins, whereby increasing lags redirect trajectories toward extinction. Therefore, similar timing-driven transitions may arise in prebiotic replicators, synthetic autocatalytic networks, and other enzyme-limited evolutionary systems. Replication timing thus emerges as a fundamental and previously unrecognized control dimension of quasispecies dynamics, with implications for evolutionary theory and antiviral intervention.

Mechanisms leading to collapse in complex ecological systems can emerge from a wide range of dynamical processes, including structural instabilities, stochastic perturbations, transient dynamics, and network interactions. Within this broader context, tipping phenomena represent a particularly important class of mechanisms behind critical transitions. The most prominent mechanism is bifurcation-induced tipping (B-tipping), typically associated with a saddle-node bifurcation. Other major mechanisms include rate-induced tipping (R-tipping), noise-induced tipping (N-tipping), shock-induced tipping (S-tipping), anomalous-noise tipping (A-tipping), phase-dependent tipping (P-tipping), and tipping at the end of long transients (LT-tipping)~\cite{Hastings2026}. In addition to these mechanisms, stochastic bifurcation tipping (SB-tipping)~\cite{Sardanyes2020,Sardanyes2024}, a mechanism distinct from N-tipping, and social tipping (SO-tipping)~\cite{Oro2023} have recently been introduced. Building on this framework, we argue that lag-time-induced tipping ($\tau$-tipping), although not directly associated with a classical bifurcation, should be recognized as a distinct tipping mechanism, as time delays alone can destabilize an otherwise stable state and drive the system toward collapse.

\vspace{0.5cm}
{\it Acknowledgements\textemdash} This research has been funded through the S. Ochoa and M. de Maeztu Program for Centers and Units of Excellence in R\&D CEX2020-001084-M (JS, EF, and JTL). FC acknowledges support from PJ93258. EF was supported by the grant PID2021-125535NB-100 (MICINN/FEDER,UE). SFE was supported by grants PID2022-136912NB-I00 funded by MCIU/AEI/10.13039/501100011033 and by “ERDF a way of making Europe”, and CIPROM/2022/59 funded by Generalitat Valenciana. JG has been supported by a fellowship from ``la Caixa'' Foundation (ID
100010434, code LCF/BQ/PR23/11980047). We thank CERCA Programme/Generalitat de Catalunya for institutional support. We also thank Esteban Domingo, Jordi García-Ojalvo, Ricard Solé, Jacobo Aguirre, Dany Bernal, and J. Tomás Lázaro for useful comments.
\bibliography{apssamp}
\vspace{-0.7cm}

\section*{Appendix}
{\it Appendix 1}\textemdash Multiple mechanisms can generate substantial time lags between genome translation and the appearance of a functional RdRp in RNA viruses. In many positive-sense RNA viruses, RdRp is produced as part of a polyprotein that requires co- and post-translational proteolytic processing before becoming active, as in picornaviruses \cite{Daijogo2011} and flaviviruses \cite{Klema2015}. Additional delays arise from folding kinetics and chaperone-assisted maturation \cite{Mine2012}, which are required for proper RdRp conformation and activity. The function of RdRp is often further dependent on post-translational modifications \cite{Wang2025} and assembly into multi-protein replication complexes, frequently associated with host's proteins \cite{Svitkin2005} and virus-induced membrane compartments \cite{Miller2011}, introducing spatial and temporal separation between synthesis and activity. In several viruses, RdRp relocalization and cooperative activation introduce additional delays before effective replication can begin, even after translation is complete \cite{Miller2011}. Finally, because the viral RNA serves simultaneously as mRNA and replication template, ribosome occupancy can transiently block RdRp access, generating an additional functional delay \cite{Daijogo2011,Gamarnik1998}; strand switching between positive and negative RNAs further contributes to temporal separation between translation and replication \cite{Daijogo2011}.

\vspace{0.1cm}

{\it Appendix 2}\textemdash The system~\eqref{eq:DIviruses} has three equilibrium points:
$P_0=(0,0,0)$, involving full extinction; 
$P_+=(x_{0+},x_{1+},p_+)$ governing genomes-RdRp persistence;
and $P_-=(x_{0-},x_{1-},p_-)$, with
$x_{0}=-2(\alpha+\beta)\varepsilon^2/\big(k(\alpha+\beta)(\varepsilon-\beta)\pm\chi\big)$,
$x_{1}=-2\gamma \mu\varepsilon^2/\big(k(\alpha+\beta)(\varepsilon-\beta)\pm\chi\big)$ and
$p=2k(\alpha+\beta)\varepsilon/\big(k(\alpha+\beta)(\varepsilon+\beta)\mp\chi\big)$
where the upper sign correspond to $P_+$ and the lower to $P_-.$ Here,
$\chi=\sqrt{-k(\alpha+\beta)\big(4\beta(\alpha+\gamma)\varepsilon^2-k(\alpha+\beta)(\varepsilon-\beta)^2\big)}$,
$\alpha=r(1-\gamma)$,
and
$\beta=\gamma(\mu-1)$. Whenever the equilibria $P_{\pm}$ are real, numerical explorations verify that, for any value of the delay parameters, $P_{0}$ and $P_{+}$ are attractors, separated by the stable manifold of the saddle equilibrium $P_{-}$.


\vspace{0.1cm}

{\it Appendix 3}\textemdash 
The equilibria of eqs.~(2) are independent of $\tau$, however, the basins are $\tau$-dependent because the basin boundary is an invariant manifold in the (history) phase space of the DDE. The separatrix dividing the basins of the origin and the persistence states, given by the stable manifold of the saddle, persists for $\tau>0$ as the stable manifold $W^s(P_-;\tau)$ in the history space, and its geometry varies with $\tau$. To see this explicitly, linearizing eq.~(2) around any nontrivial equilibrium $P_*=(x^*,y^*)$, \emph{i.e.,} $P_-$, gives the linear DDE
$\dot{\mathbf u}(t)=A_0\mathbf u(t)+A_1\mathbf u(t-\tau)$, whose eigenvalues $\lambda$ satisfy
\begin{figure}
\centering
\includegraphics[width=0.97\columnwidth]{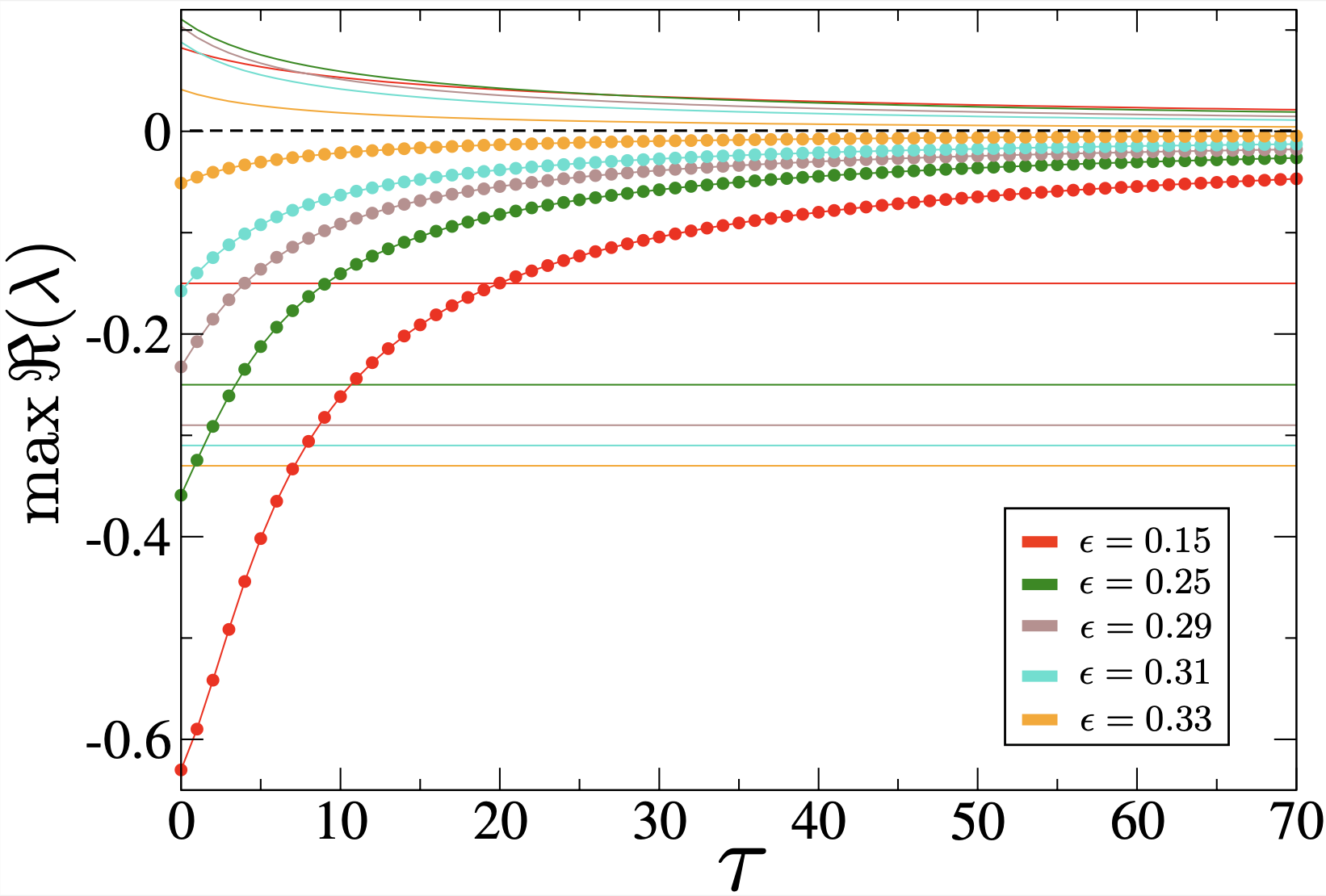}
\caption{Lag-dependent spectral stability of the equilibria of eqs.~(2) for five values of $\varepsilon<\varepsilon_c$. Shown is the rightmost real part eigenvalue, $\max\Re(\lambda)$. The equilibrium $P_0$ is $\tau$-independent and stable (horizontal lines); $P_-$ is unstable for all $\tau$, with its instability weakening at larger lags (positive curves); $P_+$ remains stable over the full delay range (dotted curves), with its attractive strength also weakening.}
\label{fig5:spectra}
\end{figure}
\begin{equation}
F(\lambda,\tau)\equiv(\lambda-a)(\lambda-d)-K\,e^{-\lambda\tau}=0,
\label{eq:char}
\end{equation}
with $a=y^*(1-2x^*)-\varepsilon$, $d=-(x^*+\varepsilon)$, and $K=x^*(1-x^*)(1-y^*)=
\varepsilon x^*(1-x^*)/(x^*+\varepsilon)$. Hence the spectrum depends on $\tau$ through the factor $e^{-\lambda\tau}$; differentiating \eqref{eq:char} yields
$\frac{d\lambda}{d\tau}=-\frac{\partial_\tau F}{\partial_\lambda F}
=-\frac{K\,\lambda\,e^{-\lambda\tau}}{2\lambda-(a+d)+K\tau e^{-\lambda\tau}},$
which is generically nonzero, showing that the (un)stable eigendirections of the saddle $P_-$ vary with $\tau$. Because the basin boundary is precisely $W^s(P_-;\tau)$, this implies that changing $\tau$ reshapes the basins of attraction of $P_0$ and $P_+$ (as also quantified numerically by the $\tau$-dependent separatrix in Fig.~4). 

\vspace{0.1cm}

{\it Appendix 4: Lag--dependent spectral stability}\textemdash 
We numerically analyzed the lag--dependent spectral stability of $P_0$, $P_+$, and $P_-$ of eqs.~(2) by solving the characteristic eq.~\eqref{eq:char} (see Appendix 3 above).  For each equilibrium and each value of $\tau$, we computed the rightmost characteristic root $\max\Re(\lambda)$, which provides a direct indicator of local spectral stability. Fig.~\ref{fig5:spectra} summarizes the results for representative values of the degradation rate $\varepsilon$. The extinction equilibrium $P_0=(0,0)$ is characterized by a double eigenvalue $\lambda=-\varepsilon$ and therefore remains locally asymptotically stable for all $\tau$. The persistence equilibrium $P_+$ remains spectrally stable for all lags, with $\max\Re(\lambda)<0$ and decreasing attraction as $\tau$ increases. In contrast, the saddle $P_-$ retains a positive real eigenvalue for all $\tau$, with weakening instability at larger delays. Thus, increasing $\tau$ induces no local bifurcations of $P_0$ or $P_+$ for the parameter values explored here; instead, it continuously deforms the stable manifold $W^{s}(P_-;\tau)$ forming the basin boundary. Temporal lags therefore reorganize basins of attraction without changing the fixed-point structure, providing the mechanism for lag-induced extinction identified in this work.

\end{document}